\documentclass[12pt]{iopart}

\usepackage{amssymb}
\usepackage{graphicx}
\usepackage{dcolumn}
\usepackage{bm}

\begin{document}
\title[Heterocyst placement strategies to maximize growth of cyanobacterial filaments]{Heterocyst placement strategies to maximize growth of cyanobacterial filaments} 

\author{Aidan I Brown and Andrew D Rutenberg}
\address{Department of Physics and Atmospheric Science, Dalhousie University, Halifax, Nova Scotia, Canada B3H 1Z9}
\ead{\mailto{aidan@dal.ca}, \mailto{andrew.rutenberg@dal.ca}}

\date{\today}

\begin{abstract}
Under conditions of limited fixed-nitrogen, some filamentous cyanobacteria develop a regular pattern of heterocyst cells that fix nitrogen for the remaining vegetative cells. We examine three different heterocyst placement strategies by quantitatively modelling filament growth while varying both external fixed-nitrogen and leakage from the filament.  We find that there is an optimum heterocyst frequency which maximizes the growth rate of the filament; the optimum frequency decreases as the external fixed-nitrogen concentration increases but increases as the leakage increases. In the presence of leakage, filaments implementing a local heterocyst placement strategy grow significantly faster  than filaments implementing random heterocyst placement strategies.   With no extracellular fixed-nitrogen, consistent with recent experimental studies of \emph{Anabaena} sp. PCC 7120, the modelled heterocyst spacing distribution using our local heterocyst placement strategy is qualitatively similar to experimentally observed patterns. As external fixed-nitrogen is increased, the spacing distribution for our local placement strategy retains the same shape while the average spacing between heterocysts continuously increases. 
\end{abstract}

\maketitle
\section{Introduction}

Cyanobacteria are prokaryotes able to grow photoautotrophically using oxygenic photosynthesis. They preferentially use ammonium or nitrate as sources of fixed-nitrogen \cite{ohmori77}. However, in conditions of low exogenous fixed-nitrogen, cyanobacteria can fix atmospheric nitrogen. Heterocystous filamentous cyanobacteria form unbranched clonal filaments of cells, and fix nitrogen within dedicated and terminally differentiated heterocyst cells \cite{flores10,kumar10}. Heterocysts are separated by clusters of photosynthetic vegetative cells that cannot  themselves produce fixed-nitrogen (fN).  Significant progress has been made in characterizing the genetic network underlying the observed heterocyst pattern \cite{flores10,golden03} that is developed in response to the absence of external fixed-nitrogen. Nevertheless, little attention has been paid to the functional role of the heterocyst pattern  itself in these model developmental organisms. 

As we shall show, a filament with too few (or inactive) heterocysts will starve of fixed-nitrogen and grow slowly without external fixed-nitrogen \cite{Jones03} while too many heterocysts, which do not grow and divide, will also inhibit growth \cite{khudyakov04}. Balancing these effects in the filament qualitatively explains the heterocyst frequency of approximately $10\%$ observed with no external fixed-nitrogen, but  does not explain the distinctive pattern of heterocyst spacings that is seen in the model cyanobacterium \emph{Anabaena} sp. PCC 7120 \cite{khudyakov04, yoon98, yoon01, toyoshima10}. Indeed, the observation that mutant strains exhibiting a distinct pattern of multiple-contiguous heterocysts (Mch) show reduced growth \cite{callahan01} indicates that heterocyst placement is important. In this paper, we use quantitative modelling to explore the hypotheses that simple heterocyst placement strategies can affect filament growth, and that observed heterocyst patterns reflect placement strategies that maximize growth. 

Previous models of cyanobacterial spacings have directly compared model and experimental distributions.  Meeks and Elhai \cite{meeks02} compared the expected heterocyst spacing for randomly spaced heterocysts to the experimentally measured distribution and found clear disagreement. Wolk and Quine \cite{wolk75} examined a diffusible inhibitor mechanism in which a radius of inhibition around existing heterocysts depends on diffusion and degradation of the inhibitor, and obtained qualitative agreement with early observed spacing distributions for {\em Anabaena} sp. PCC 7120 (hereafter simply PCC 7120).  These models identify  lateral inhibition as a plausible mechanism behind the observed heterocyst patterns but could not examine why (or if) the observed pattern is optimal, or how the pattern may change under different experimental conditions. 

We hypothesize that leakage of fN from the cyanobacterial filament may distinguish between heterocyst patterns, in terms of growth rates.  Evidence for leakage was first reported by Fogg \emph{et al} \cite{fogg49,walsby75}, who found fN products outside the filament. Supporting this,  Paerl \cite{paerl78} observed bacteria clustering around cyanobacterial filaments, particularly at junctions between heterocysts and vegetative cells, and indicated a possible symbiosis based on enhanced nitrogen fixation and leaked fixation products. Thiel \cite{thiel90}  found protein proteolysis byproducts in the extracellular medium after fixed-nitrogen starvation.   Significant leakage is also consistent with the observation of reduced diazotrophic growth in strains with impaired amino acid uptake transporters \cite{pernil08,picossi05}. Nevertheless, we are not aware of any  quantitative measurements of the rate of leakage. Qualitatively, a  regular pattern of heterocysts would minimize the distance traveled by fN products so as to minimize leakage from vegetative cells, leaving more fN available for growth.

In this paper, we examine heterocyst frequencies that maximize filament growth within the context of a quantitative transport model \cite{brown12} that incorporates fixed-nitrogen transport, vegetative cell growth, and fixed-nitrogen production at heterocysts.  We explore the impact on growth of different heterocyst placement strategies, including random placement, and find that they are almost indistinguishable in the absence of leakage of fN from the filament, but clearly distinct with leakage.  We find that the  heterocyst spacing patterns corresponding to maximal filament growth are qualitatively similar to those seen experimentally.

The genera \emph{Anabaena}, which includes heterocystous cyanobacteria, is widely distributed geographically in freshwater lakes \cite{gibson82, wetzel75} where fixed-nitrogen is one of the major substances limiting growth \cite{lund65, nielsen60, reynolds75}. Nitrification turns ammonia into nitrite and then nitrate \cite{wetzel75, stumm96}. Nitrate levels in lakes range from 0-10 mg/l (0-10$\times$10$^{22}$ m$^{-3}$) in unpolluted freshwater, but vary both seasonally and spatially \cite{wetzel75, downing92}.  Heterocystous cyanobacteria are also found in the oceans \cite{charpy10,stacey77} where nitrate levels have been recorded to vary from approximately 0-50 $\mu$mol/kg \cite{tyrrell99} (0-3$\times$10$^{22}$ m$^{-3}$). It has long been known that sources of fixed-nitrogen initially present in the medium can increase the mean spacing between heterocysts \cite{fogg49,ogawa69} and that heterocysts will differentiate at non-zero levels of external fixed-nitrogen (efN) \cite{fogg49}.  Field studies of heterocystous cyanobacteria also show both a range of heterocyst counts and a range of efN concentrations, with significant positive correlations between the two (see e.g. \cite{horne79}).   In addition, steady-state chemostat experiments show that {\em Anabaena flos-aquae} can adjust nitrogen fixation to achieve constant growth, within approximately 10\%, as efN concentrations are varied \cite{elder84}.  While growth independence from efN seems desirable in the face of  environmental variability, it raises the question of how and how well it is achieved in terms of heterocyst fraction and pattern.    

The heterocyst pattern changes with time as it evolves towards a steady-state distribution after efN deprivation \cite{khudyakov04, yoon01, toyoshima10}.  Earlier models \cite{meeks02, wolk75} have focused on the early heterocyst pattern observed 24 h after efN deprivation. While the early pattern and the later patterns are qualitatively similar, with broad distributions of heterocyst spacings ranging from zero to more than twenty cells between heterocysts with a peak at about a ten cell spacing, we focus on the steady-state distribution in this paper.  We explore the hypothesis  that cyanobacterial filaments use a unified heterocyst placement strategy even at non-zero levels of efN.   With all of our strategies, we find that maximal growth is observed with heterocyst frequencies that decrease continuously with increasing levels of efN. 

\section{Model}
\label{sec:model}

\subsection{Fixed-nitrogen Transport}

Our model of fixed-nitrogen (fN) transport and incorporation is adapted from Brown and Rutenberg \cite{brown12}, with the addition of import from the extracellular medium. That study found that periplasmic transport was not required to explain nanoSIMS studies of fixed-nitrogen profiles \cite{popa07}, so we do not include periplasmic transport in our model.  Our model tracks the total amount of freely-diffusing fN, $N(i,t)$, in each cell $i$ vs. time $t$: 
\begin{equation}
	\label{eq:cytonitrogen}
	\frac{d}{dt}N(i,t)=\Phi_{tot}(i,t)+D_I \rho_{efN} l(i,t)-D_LN(i,t)+G_i,
\end{equation}
where $\Phi_{tot}(i) \equiv \Phi_{R}(i-1)+\Phi_{L}(i+1)-\Phi_{L}(i)-\Phi_{R}(i)$ is the net diffusive flux into cell $i$, equal to the sum of the incoming fluxes minus the outgoing fluxes.  Each cell has two outgoing fluxes, $\Phi_L$  and $\Phi_R$ to its left and right neighbours, respectively, and two incoming fluxes from its neighbours. Following Fick's law, each flux is the product of the local density $N(i,t)/l(i,t)$ and a transport coefficient $D_C$. We use $D_C$=1.54 $\mu$m s$^{-1}$ between two vegetative cells and $D_C$=0.19 $\mu$m s$^{-1}$ between a vegetative cell and a heterocyst \cite{brown12} --- we keep these values fixed in this paper.

\begin{figure}[!ht]
 \begin{center}
 \includegraphics[width=3.25in]{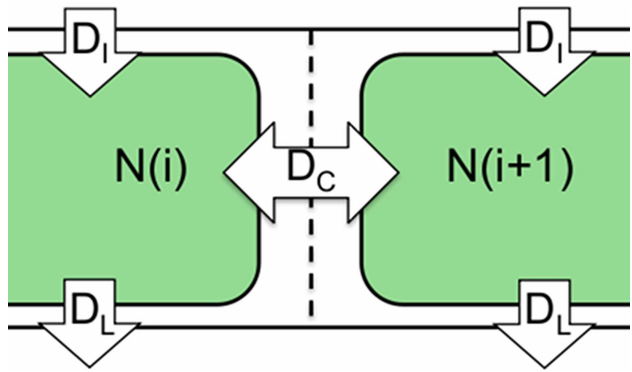}
  \end{center}
  \caption{\label{fig:first}Schematic of fixed-nitrogen transport as represented by Eqn.~\ref{eq:cytonitrogen}.  $N(i)$ is the amount of freely-diffusing fixed-nitrogen for  cell $i$. $D_C$ governs transport between cells, $D_I$ governs import from the external medium into the cell, and $D_L$ controls loss from the cell to the external medium.}
\end{figure}

In addition to fluxes along the filament, $D_I$ is the coefficient for import from the external medium with efN concentration $\rho_{efN}$ to the cell of length $l_i$ and $D_L$ is the coefficient for losses from the cell to outside the filament. These transport processes are shown schematically in Fig.~\ref{fig:first}.

To estimate $D_I$, we use Fogg's observations of heterocyst frequency in a medium to which a fixed amount of efN in the form of ammonia was added \cite{fogg49}. Heterocyst frequency decreased in time due to vegetative growth until the ammonia concentration dropped to $4 \times 10^{-5}$M (see p. 245 of \cite{fogg49}). We assume that a slightly higher concentration of efN, $5 \times 10^{-5}$M ($\rho_{efN}$=3$\times$10$^{22}$m$^{-3}$), is just sufficient for maximal growth. From Eqn.~\ref{eq:cytonitrogen} the amount of fixed-nitrogen imported from the external medium in a time $T$ is $D_I\rho_{efN}l T$. For a cell to double in length over a period of time $T$=20h it needs approximately 1.4$\times$10$^{10}$ N atoms to be imported \cite{brown12}. Using $l$=$l_{min}$=2.25$\mu$m, this yields $D_I =2.9 \times 10^{-18}$m$^3/ (\mu$m$\cdot$s).

We also include a loss term with coefficient $D_L$ in Eq.~\ref{eq:cytonitrogen}, following \cite{fogg49, walsby75, paerl78, thiel90, pernil08, picossi05}.  Loss is expected to be proportional to the cytoplasmic density, $N/(A l)$, where $l$ is the length of the cell and $A$ is the cross-sectional area, and also to the number of transporters, which will themselves be proportional to the cell length. This results in a loss term that is independent of length $l$, with a coefficient $D_L$ with units of s$^{-1}$. There are no direct measurements of leakage rates of fN from cyanobacterial filaments that we are aware of. We take the loss rate as small compared to the import rate, and so use either $D_L$=0.01$D_I/A$ (referred to as 1$\%$ loss) and $D_L$=0.1$D_I/A$ (referred to as 10$\%$ loss). We use a fixed $A= \pi \mu$m$^2$ corresponding to a radius of $1 \mu$m.

In addition to transport terms, there is  also a source/sink term $G_i$ in Eqn.~\ref{eq:cytonitrogen} that describes fN production and consumption in the heterocysts and vegetative cells, respectively. This $G$ term is discussed in the next section.

\subsection{Cell Growth and Division}
\label{sec:growth}
Following \cite{brown12}, we take PCC 7120 cells to have a minimum size of $l_{min}$ = 2.25$\mu$m and a maximum size of $l_{max}=2 l_{min}$ \cite{flores10,kumar10}.  When a cell reaches $l_{max}$ it is divided into two cells of equal length, each of which is randomly assigned a new growth rate and half of the fixed-nitrogen in the parent cell. We initialized lengths randomly from an analytical steady-state distribution of cell lengths  ranging between $l_{min}$ and $l_{max}$ \cite{powell56} and used an average doubling time $T_D = 20$h \cite{picossi05,herrero90}. We define a minimum doubling time $T_{min}$ = $T_D-\Delta$ and a maximum doubling time $T_{max}$ = $T_D+\Delta$. A doubling time $T$ is randomly and uniformly selected from this range and converted to a growth rate $R$=$l_{min}/T$.  We take $\Delta=4.5$h \cite{brown12, allard07}.

In Eqn.~\ref{eq:cytonitrogen}, for heterocysts $G_i=G_{het}$.   The heterocyst fN production rate, $G_{het}= 3.15 \times 10^6$s$^{-1}$, is chosen to supply the growth of approximately 20 vegetative cells. 

In Eqn.~\ref{eq:cytonitrogen},  for vegetative cells $G_i=G_{veg}$ is a sink term determining the incorporation rate of cytoplasmic fN removed to support cellular growth.  $G_{veg}$ depends on the actual growth rate $R$ of the cell (in $\mu$m/s), which in turn depends upon the locally available cytoplasmic fN: 
\begin{equation}
	\label{eq:gequation}
	G_{veg}=-g R(R_i^{opt},N(i,t)),
\end{equation}
where $g$ is the amount of fixed-nitrogen a cell needs to grow per $\mu$m; $g = 1.4 \times 10^{10}/l_{min} \simeq  6.2 \times10^9 \mu$m$^{-1}$ \cite{brown12}.  We assume that cells increase their length at their maximal growth rate $R_i^{opt}$ as long as there is available cytoplasmic fixed-nitrogen. Otherwise, they can only grow using the fixed-nitrogen flux into the cell from neighbouring cells:
\begin{equation}
\label{eq:requation}
R=
   \cases{
      R_i^{opt}, & if $N(i,t)>0$\cr
      min(\Phi_{in}/g, R_i^{opt}) & if $N(i,t)=0$.\cr
   }
\end{equation}
Note that cells with $N=0$ may still grow, but will be limited by the incoming fluxes of fixed-nitrogen from both adjoining cells and the external medium, $G_{veg}=\Phi_{in} = \Phi_R(i-1) + \Phi_L(i+1) + D_I\rho_{efN} l(i)$.  

\subsection{Heterocyst Placement}

While we know that experimentally observed heterocyst patterns are not due to random placement \cite{meeks02}, it is useful to evaluate the  steady-state growth rate achievable with random heterocyst placement. We will use random placement as a point of reference for other heterocyst placement strategies. We consider three simple strategies: random placement, random placement with no contiguous heterocysts, and local placement. For all of them, once a heterocyst is placed it immediately stops growing ($R_i^{opt}=0$) and fixes nitrogen ($G_i=G_{het}$). Commitment to differentiation does not occur until after approximately 8 h of efN deprivation and can take as long as 14 h \cite{yoon01, ehira03, thiel01}. Heterocysts mature and begin to fix nitrogen approximately 18-24 h after efN deprivation \cite{kumar10, yoon01, ehira03} (though see \cite{allard07}). For random strategies, heterocyst placement corresponds to when heterocysts begin to produce fN. For local placement, we have an explicit delay (see below).

Our first, reference, strategy is random placement (``random''). The heterocyst fraction $f$ is fixed, and during filament growth a random vegetative cell is replaced with a heterocyst whenever possible but without exceeding $f$. 

Our second strategy (``no-Mch'') reflects the observation that contiguous heterocysts are not observed during normal development \cite{flores10,kumar10}.  It consists of our random strategy, but with the additional restriction that vegetative cells adjacent to existing heterocysts are never  selected for development.

Our third strategy is local heterocyst placement (``local''). Any vegetative cell that has been continuously starving for a defined interval $\tau$ is changed into a heterocyst.  Starvation is defined as $N(i)=0$, so that these cells have reduced growth $R < R^{opt}$ for a significant period of time. We vary $\tau$ within the range $1-20$ h. Starvation occurs due to distance from heterocysts \cite{brown12}, but can also reflect local clusters of fast-growing cells.

\subsection{Some numerical details}
Periodic boundary conditions were used to minimize end effects. Filaments were initiated with no heterocysts and the different strategies were followed until a steady-state was reached. For the local strategy, a ``no-Mch'' rule was followed for the first 24 hours of differentiation to reduce initial transients. 

The growth rate constant $\mu$ was calculated every six simulated hours using the total length of the filament $L(t)$, where $L(t)=L(t-6h)e^{\mu \cdot 6h}$. Heterocyst frequency was sampled every six simulated hours by dividing the number of heterocysts by the total number of cells in the filament. Heterocyst spacings were also recorded every six simulated hours.  Measurements were averaged daily, and the results of ten independently seeded runs were used to determine an overall daily average and standard deviation.  All data shown is for the fifth day, which exhibits steady-state for the parameters explored (in comparison with data from the fourth day).

We begin each simulation with 100 cells. Most of the phenomena we investigate occurs in filaments with a growth rate constant greater than $\mu=$0.02 h$^{-1}$, which would allow the filament to grow to more than 1100 cells after five days. We use a computational timestep $\Delta t=$0.01s with a simpler Euler discretization of Eqn.~\ref{eq:cytonitrogen}; smaller timesteps yield indistinguishable results.

For random heterocyst placement strategies, optimal heterocyst frequencies for growth without leakage in Fig.~\ref{fig:second}(c), and the corresponding growth rates in Fig.~\ref{fig:second}(d), were found by locally using the standard Marquardt-Levenberg fit algorithm with a quadratic function to the left of the optimum frequency and a linear function to the right, with the two functions meeting at the optimum frequency. Optimum heterocyst frequencies with leakage in Figs.~\ref{fig:fourth}(b) and (d), and the corresponding growth rates in Figs.~\ref{fig:fourth}(a) and (c), were found by a least squares quadratic fit near the maximal growth rate. 

\section{Results}
\label{sec:results}

\subsection{No Leakage}

\begin{figure}[!ht]
 \begin{center}
  \begin{tabular}{cc}
    \includegraphics[width=3.0in]{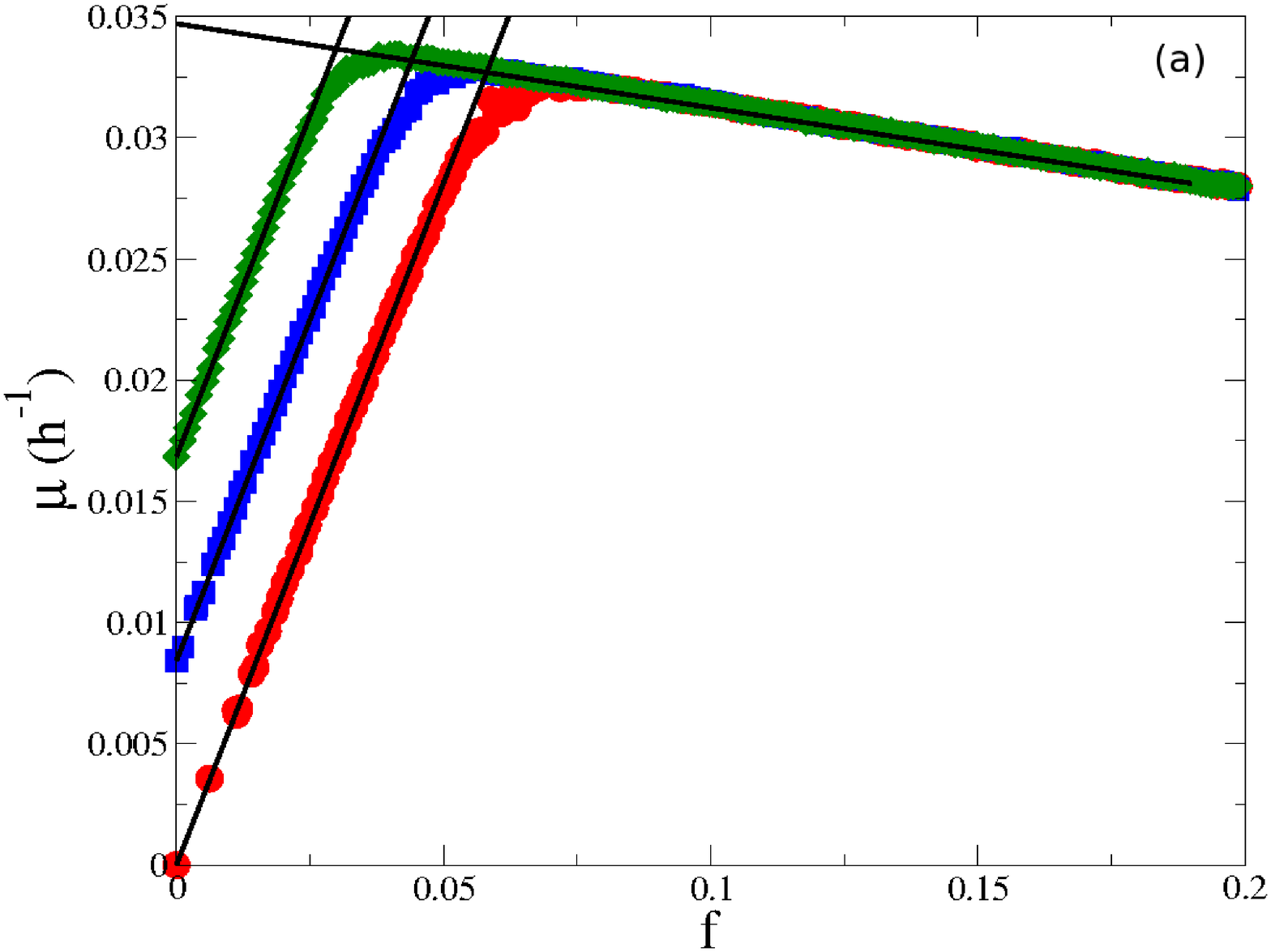}&\includegraphics[width=3.0in]{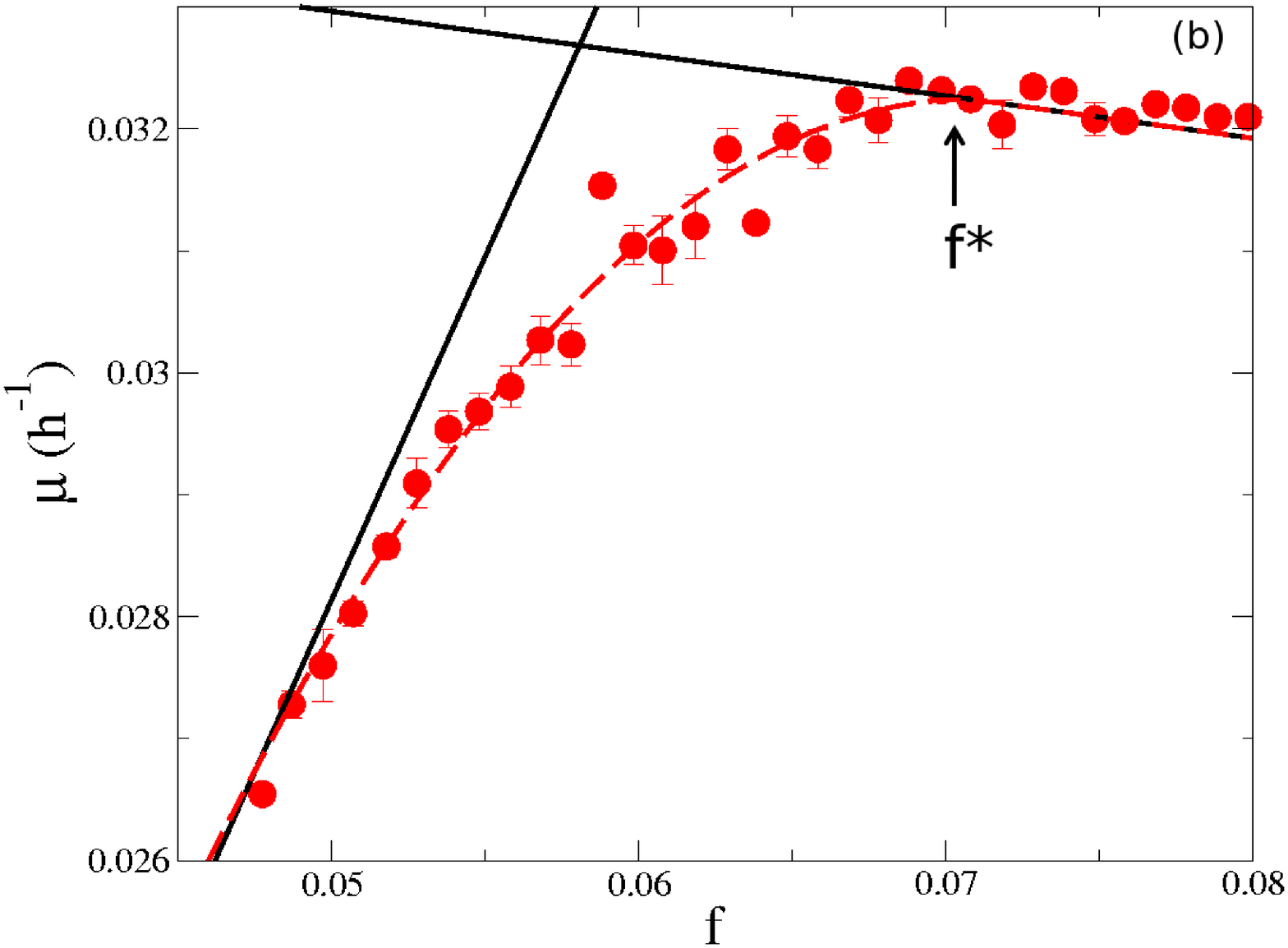} \\
    \includegraphics[width=3.0in]{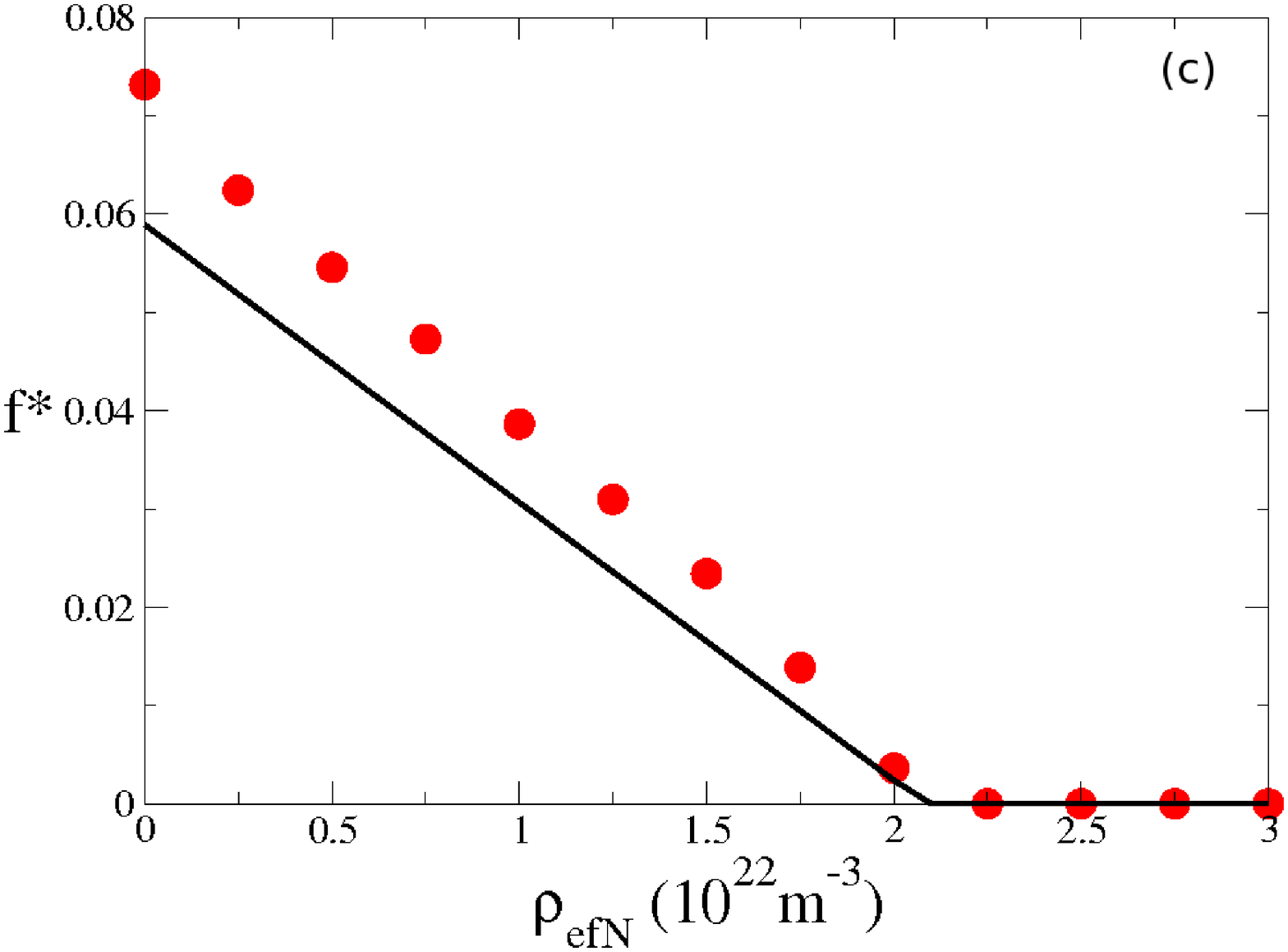}&\includegraphics[width=3.0in]{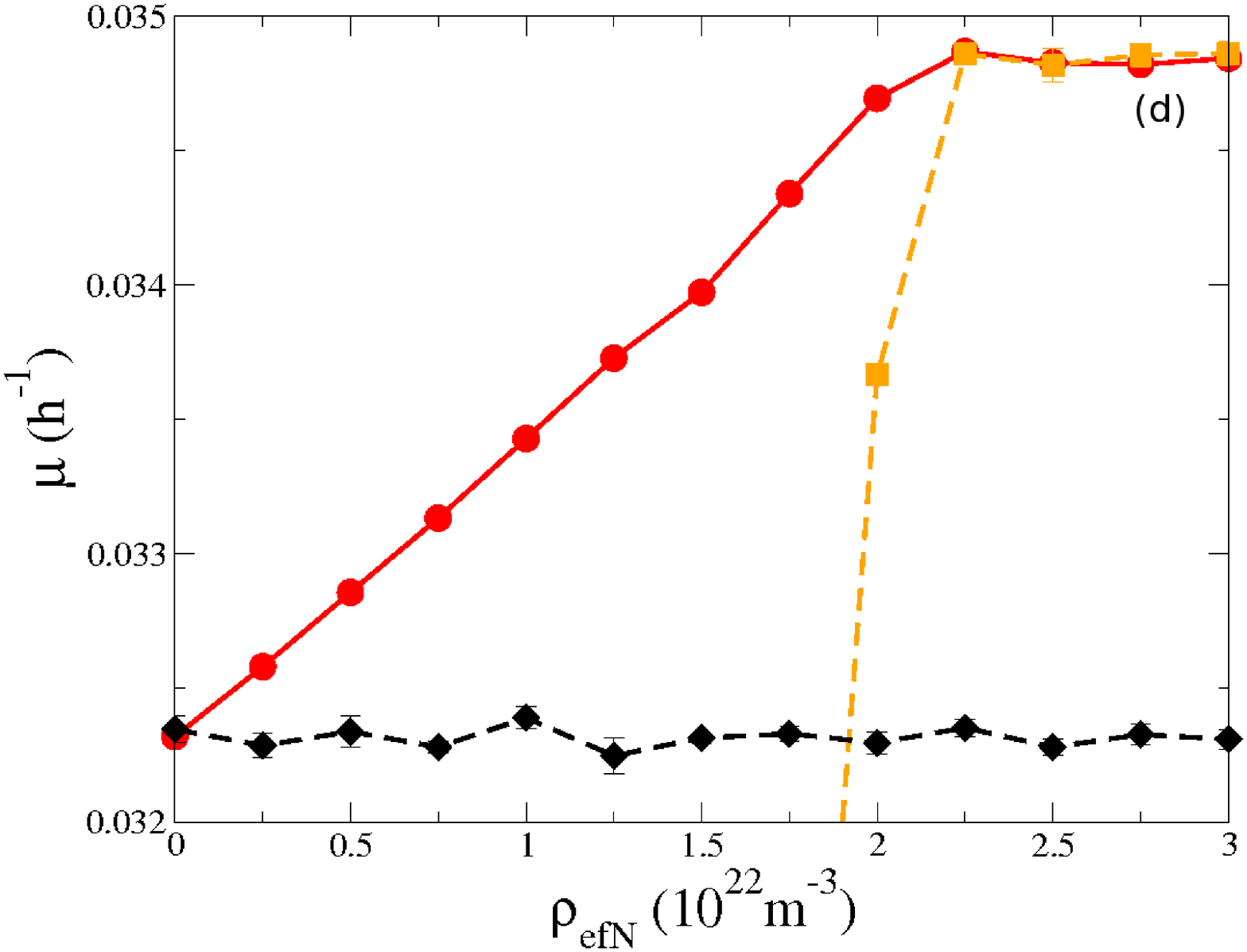}
  \end{tabular} 
  \end{center}
  \caption{\label{fig:second}(a) growth rate exponent $\mu$ vs. heterocyst frequency $f$ for systems with zero leakage ($D_L=0$) and random heterocyst placement. Red circles indicate $\rho_{efN}=0$, blue squares indicate $\rho_{efN} = 0.5 \times 10^{22} m^{-3}$, and green diamonds indicate $\rho_{efN}= 1 \times 10^{22} m^{-3}$. Black lines indicate asymptotic behavior from Eqns.~\ref{eq:starving} and \ref{eq:excess}. (b) highlights the data near the maximal $\mu$ in (a).  The red dashed curve indicates the best fit near the indicated optimal heterocyst fraction $f^\ast$. (c) optimum heterocyst frequency $f^*$ vs. $\rho_{efN}$  using random heterocyst placement with zero leakage. Numerical data is shown as red circles while the frequency described by the intersection of Eqns.~\ref{eq:starving} and \ref{eq:excess} is shown as a solid black line. (d) growth rate $\mu$ vs. $\rho_{efN}$. Red circles with a solid line indicate $\mu^\ast$ from random heterocyst placement with zero leakage. Orange squares with a dashed orange line indicate $\mu$ for a filament with no heterocysts, while the black diamonds and black dashed line indicate $\mu$ for a filament maintaining a constant heterocyst frequency that is optimal at $\rho_{efN}=0$.}
\end{figure}

We first examined systems with randomly placed heterocysts in filaments with zero leakage. Qualitatively there are two growth regimes: starving or excess fixed-nitrogen. At small heterocyst fractions and small levels of efN,  starved growth will be determined by the amount of fN produced by heterocysts as well as fN imported from outside the filament. The fN from heterocysts is proportional to their fraction $f$, while the fN imported from outside the filament is proportional to $\rho_{efN}$:
\begin{equation}
\label{eq:starving}
	\mu_{starve} =  \frac{G_{het}}{g\cdot l_{avg}}f+\frac{D_I}{g}\rho_{efN},
\end{equation}
where $l_{avg}$ is the length of an average cell and $g$ is the amount of fN per unit length needed for growth.  At large heterocyst fractions or with large levels of efN, sufficient fN is present for growth but only the vegetative cells, with fraction $1-f$, will grow: 
\begin{equation}
\label{eq:excess}
	\mu_{excess} =  \frac{R\cdot ln(2)}{l_{min}}(1-f),
\end{equation}
where the $ln(2)$ factor is needed to convert the growth rate $R$ of a single cell to a growth rate exponent $\mu$. Both of these limiting behaviors $\mu_{starve}$ and $\mu_{excess}$ are shown as straight black lines in Figs.~\ref{fig:second}(a) and (b). 

The modelling results for filaments with different levels of efN and with zero leakage ($D_L=0$), shown in Fig.~\ref{fig:second} (a), agree well with the limiting behaviours for small or large $f$.  There is a clear optimal heterocyst frequency, $f^\ast$, at which growth is maximal. Increasing the amount of efN shifts $\mu_{starve}$ up, while $\mu_{excess}$ is unaffected. As efN increases, the optimal frequency decreases and the corresponding maximal growth rate $\mu^\ast$ increases.

The modelled growth rates are somewhat below the limiting regimes near $f^\ast$, as highlighted in Fig.~\ref{fig:second}(b).  This growth deficit appears to be due to finite diffusivity of fN within the filament, which limits the reach of excess fN from regions with excess heterocysts in the face of continued expansion of vegetative regions due to ongoing growth. Indeed, the growth deficit disappears when $D_C \rightarrow \infty$ (data not shown).  The growth deficit is also smaller when heterocysts are placed close to starving cells (see below). We do not yet have an analytical treatment of this growth deficit, though it is intriguing. As a result of the growth deficit, the optimal heterocyst frequency from our quantitative model, $f^\ast$, is larger than given by the intersection of $\mu_{starve}$ and $\mu_{excess}$, as shown in Fig.~\ref{fig:second}(c).

The maximal growth rates, $\mu^\ast$, corresponding to the optimal heterocyst frequencies $f^\ast$, are plotted with solid red lines vs. $\rho_{efN}$   in Fig.~\ref{fig:second}(d).  The dashed orange lines show the growth rate with no heterocysts ($f=0$), which exhibits sharply reduced growth at smaller $\rho_{efN}$. The dashed black line shows the growth rate exhibited by the fixed heterocyst fraction $f_0$ that is optimal for $\rho_{efN}=0$, which exhibits a constant but reduced growth when $\rho_{efN}>0$.  In general, filaments that maintain an optimal heterocyst fraction by maintaining $f=f^\ast$ as $\rho_{efN}$ varies will outgrow filaments with any given fixed heterocyst fraction. 

\subsection{Leakage}

\begin{figure}[!ht]
 \begin{center}
    \includegraphics[width=3.25in]{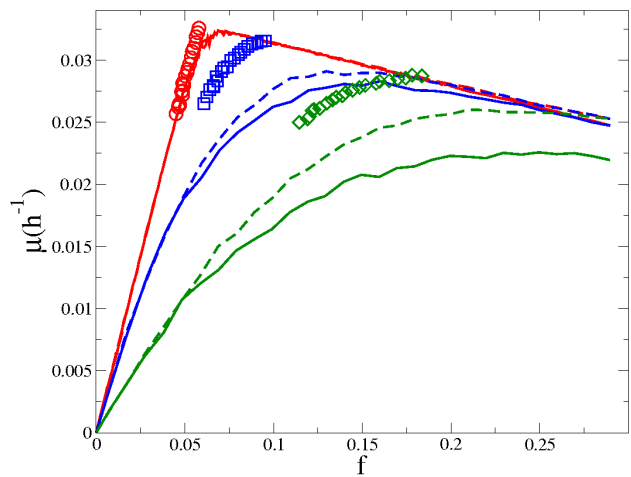}
  \end{center}
  \caption{\label{fig:third}Growth rate constant $\mu$ vs. heterocyst frequency $f$ for different heterocyst placement strategies and leakage levels with zero external fixed-nitrogen. Zero leakage is shown in red, 1$\%$ leakage is shown in blue, and 10$\%$ leakage is shown in green. Solid lines are random heterocyst placement, dashed lines are no-Mch heterocyst placement.  Red circles, blue squares, and green diamonds show data for local heterocyst placement for zero, 1$\%$, and 10$\%$ leakage respectively. The local heterocyst placement data points show variation of the period of starvation until commitment $\tau$ from 1-20 h, with shorter $\tau$ exhibiting larger $\mu$.}
\end{figure}

As shown in Fig.~\ref{fig:third}, with no leakage the random (solid red line) and no-Mch (dashed red line) heterocyst placement strategies have almost indistinguishable growth rates. Local heterocyst placement (red circles) leads to slightly faster growth with short wait times $\tau=1$ h, but shows significantly slower growth at longer wait times comparable to heterocyst maturation times. Note that $f$ is not independently controlled with local placement, so that only a narrow range of $f$ is seen as $\tau$ is varied.  Faster local growth is seen with smaller $\tau$ as starving cells are provided with fN earlier, and experience shorter periods of restricted growth. 

With non-zero leakage, the three heterocyst placement strategies produce noticeably different growth curves. At 1$\%$ leakage (blue lines and squares), the growth rates with no-Mch are slightly above the random placement strategy. More striking is the dramatic improvement with local placement, which has significantly better growth at any given heterocyst frequency, but also better maximal growth $\mu^\ast$ for $\tau \in [1,13]$h. The same trends continue when leakage is increased to 10$\%$ (green lines and green diamonds): the  growth rates of filaments with local heterocyst placement are higher than the maximal growth rates of other strategies for $\tau \in [1,17] $h, and we also see that the corresponding heterocyst frequencies $f^\ast$ are much lower with local strategies than with either random placement or no-Mch placement strategies. 

Note that $f^\ast$ and $\mu^\ast$ correspond to the heterocyst frequency and growth rate exhibited by the local placement strategy for a given delay time $\tau$, or to the {\em growth-optimized} heterocyst frequency and corresponding growth rate exhibited by the random or no-Mch placement strategies. 

In general, leakage of fN from the cyanobacterial filament will more strongly inhibit growth if heterocysts are not placed close to starving vegetative cells. In any case, leakage decreases the growth rate at a given heterocyst frequency, and so leads to a larger heterocyst frequency $f^\ast$ and correspondingly decreased growth rate $\mu^\ast$ with both the local strategy and the growth-optimized random strategies. 

\subsection{Varying External Fixed-nitrogen Concentration}

\begin{figure}[!ht]
 \begin{center}
  \begin{tabular}{cc}
    \includegraphics[width=3.0in]{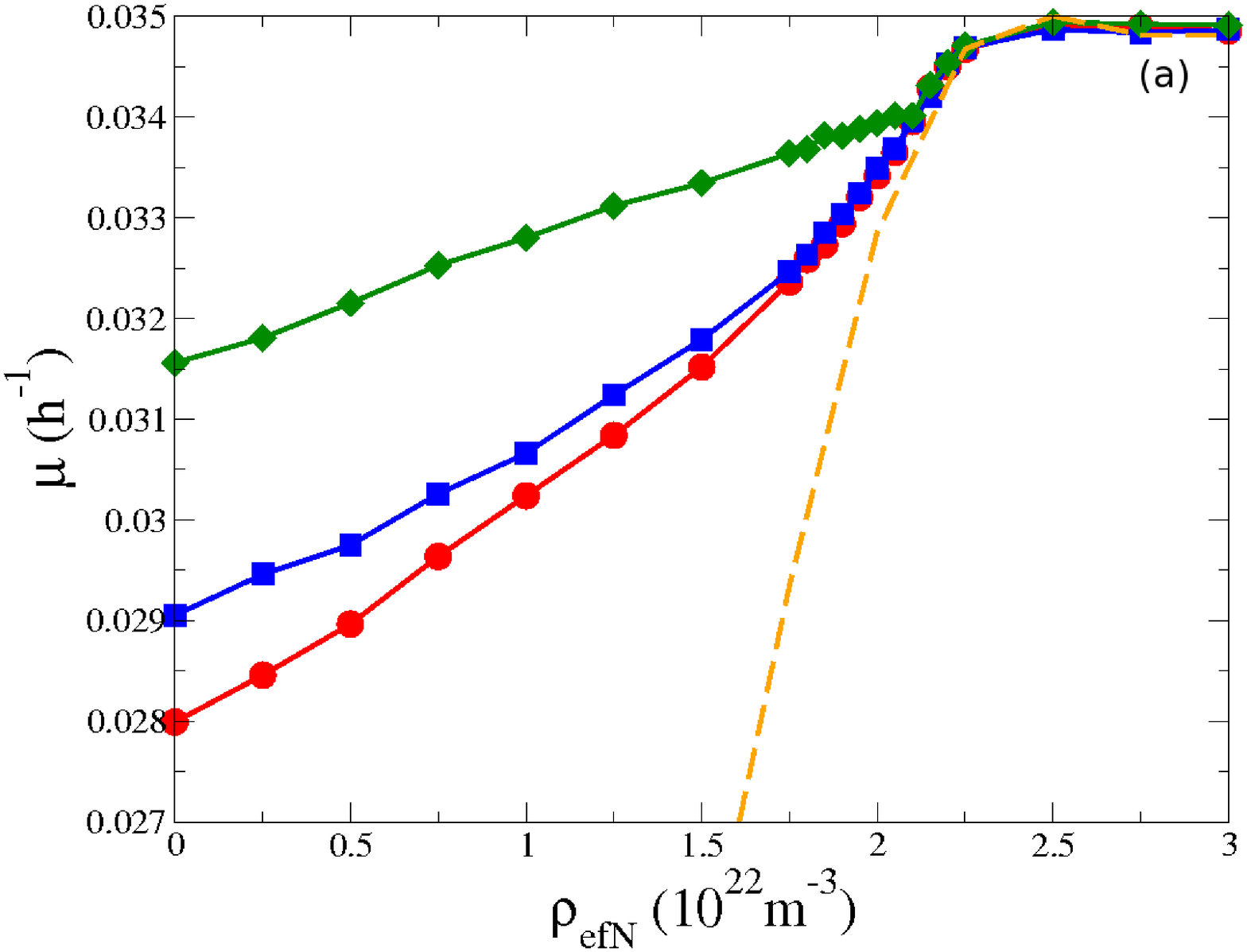}&\includegraphics[width=3.0in]{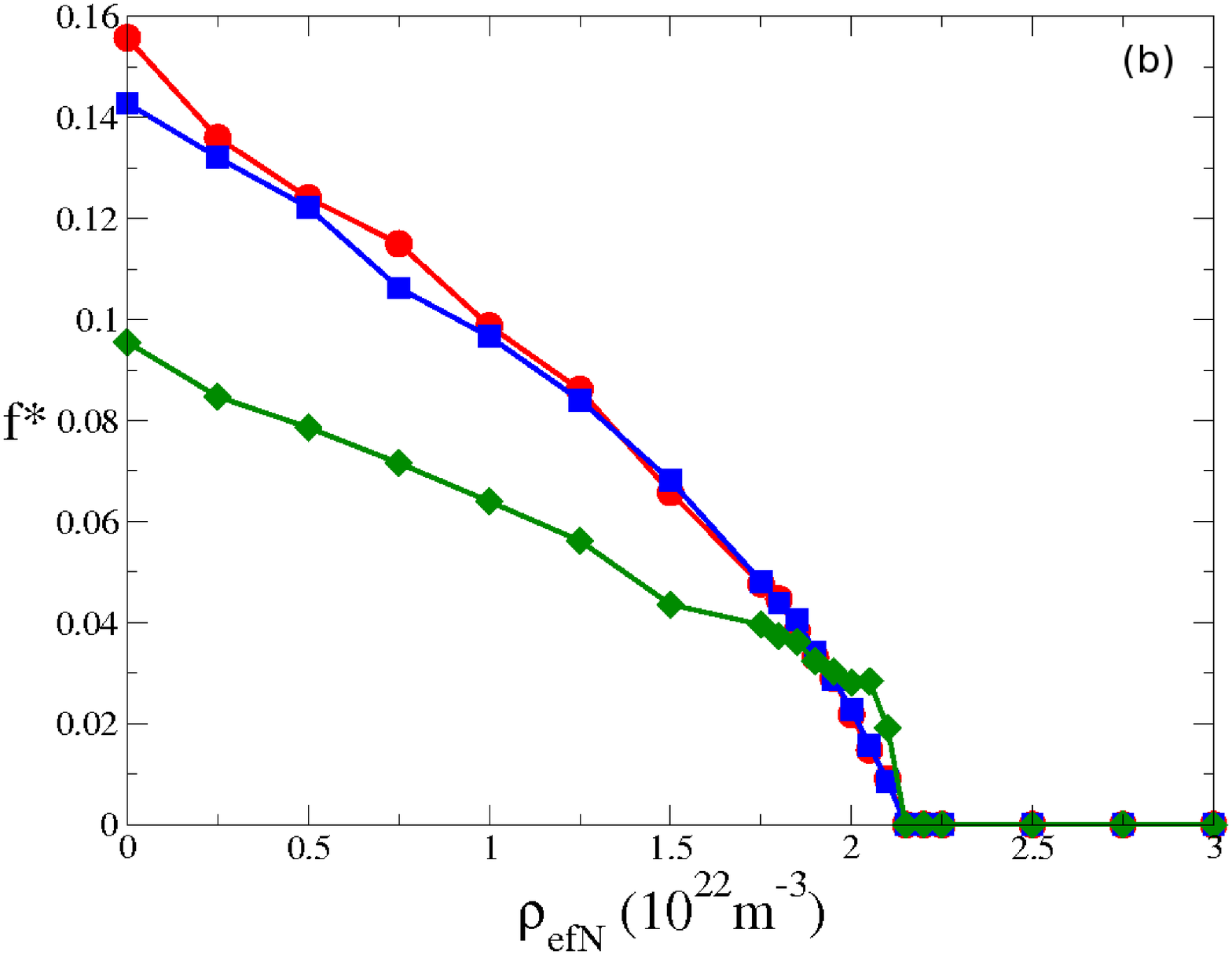} \\
    \includegraphics[width=3.0in]{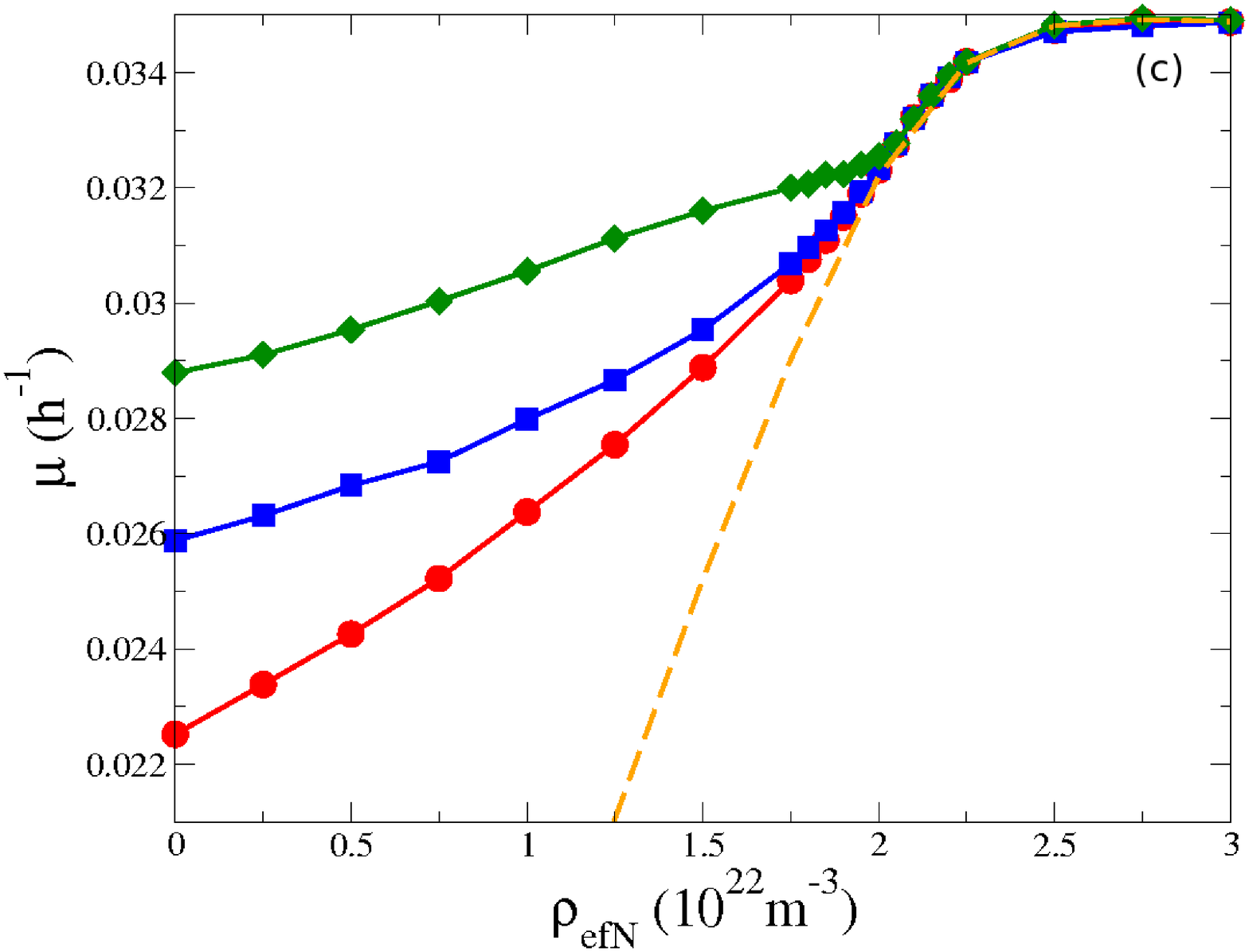}&\includegraphics[width=3.0in]{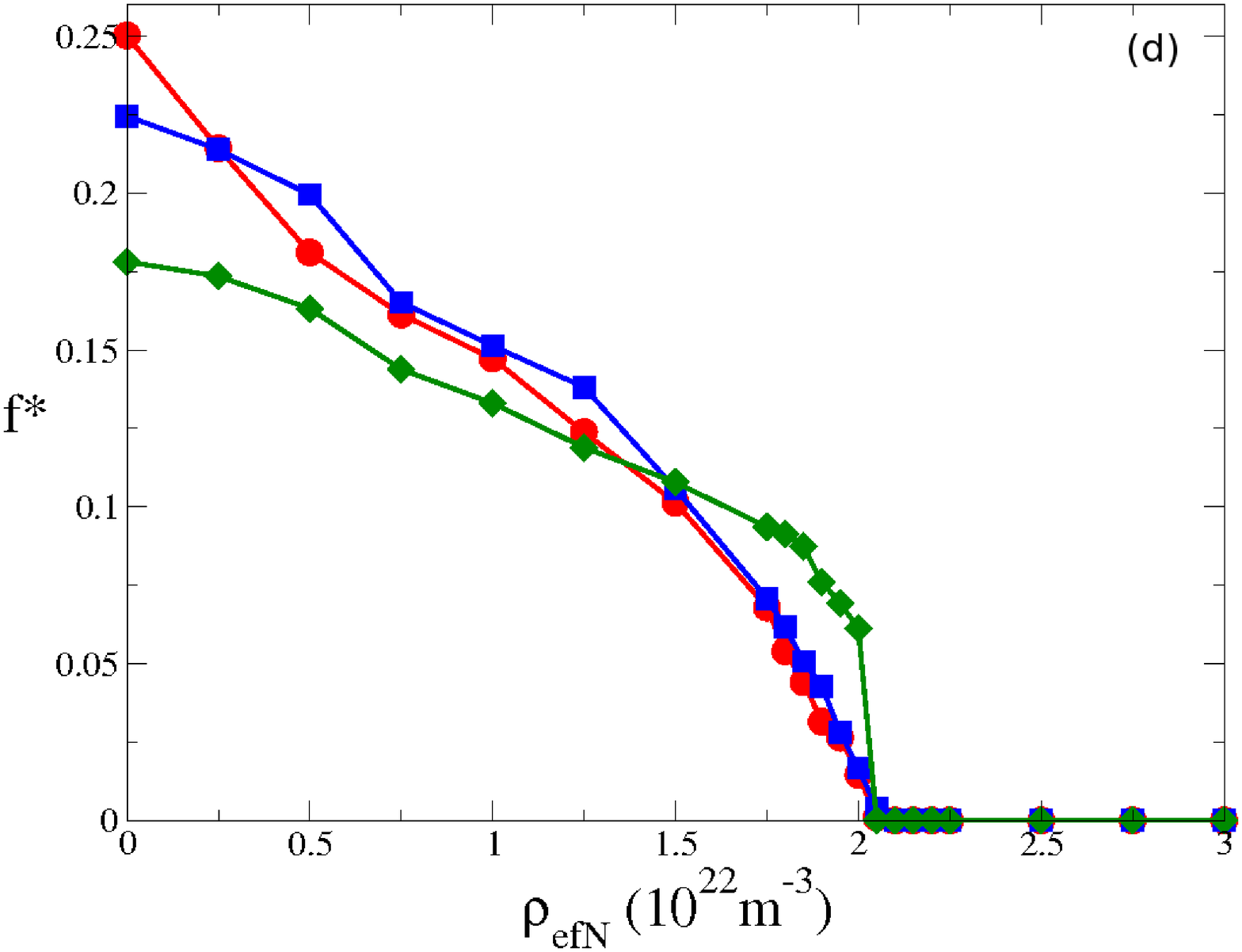}
  \end{tabular} 
  \end{center}
  \caption{\label{fig:fourth}(a) growth rate constant $\mu$ vs. external fixed-nitrogen concentration $\rho_{efN}$ for filaments with 1$\%$ leakage. Red circles with a solid line are random heterocyst placement at the optimum frequency, blue squares with a solid line are no-Mch heterocyst placement at the optimum frequency, green diamonds with a solid line are local heterocyst placement, and the orange dotted line is filaments with no heterocysts. (b) heterocyst frequency at which maximum growth occurs vs. external fixed-nitrogen concentration $\rho_{efN}$. Red circles with solid line are random heterocyst placement, blue squares with solid line are no-Mch heterocyst placement and green diamonds with solid line are local heterocyst placement. (c) and (d) are similar runs labeled in the same fashion as (a) and (b), respectively, with higher leakage of 10$\%$.}
\end{figure}

When the external fixed-nitrogen concentration, $\rho_{efN}$, is increased from zero the differences between the heterocyst placement strategies diminish. Fig.~\ref{fig:fourth}(a) shows growth, $\mu^\ast$,  vs. $\rho_{efN}$ for all three placement strategies with 1$\%$ leakage. Also shown with the orange-dashed line is the growth expected with no heterocysts ($f=0$). We see that local heterocyst placement has a significantly higher growth rate than  optimized random or no-Mch strategies. The differences are largest at $\rho_{efN}=0$, decrease as efN increases, and vanish when all growth is supported by efN alone.  Similar qualitative behavior is seen with 10$\%$ leakage, in  Fig.~\ref{fig:fourth}(c), though the growth rates are somewhat lower.   

Figs.~\ref{fig:fourth}(b) and (d) show the heterocyst frequency, $f^\ast$, for all three heterocyst placement strategies for a range of efN, for 1$\%$ and 10$\%$ leakage respectively. For most of the lower fixed-nitrogen levels, local heterocyst placement has a significantly lower heterocyst frequency than optimized random or no-Mch heterocyst placement. When heterocysts are placed near starving cells fewer heterocysts are necessary to satisfy the fixed-nitrogen requirements for the growing vegetative cells.   As $f^\ast$ approaches zero, the trend reverses. The local heterocyst placement has a higher $f^\ast$ than either random or no-Mch placement because local can continue to beneficially place heterocysts near starving cells as $f$ increases and fewer cells starve, while random strategies cannot.  The reversal of this trend is particularly apparent with larger leakage levels (Fig.~\ref{fig:fourth}(d)), where placement of heterocysts close to fast-growing vegetative cells is particularly important, and leads to a noticeable non-linearity of the heterocyst frequency $f^\ast$ with local placement at larger $\rho_{efN}$.  

\subsection{Heterocyst Spacing}

\begin{figure}[!ht]
 \begin{center}
  \begin{tabular}{cc}
    \includegraphics[width=3.0in]{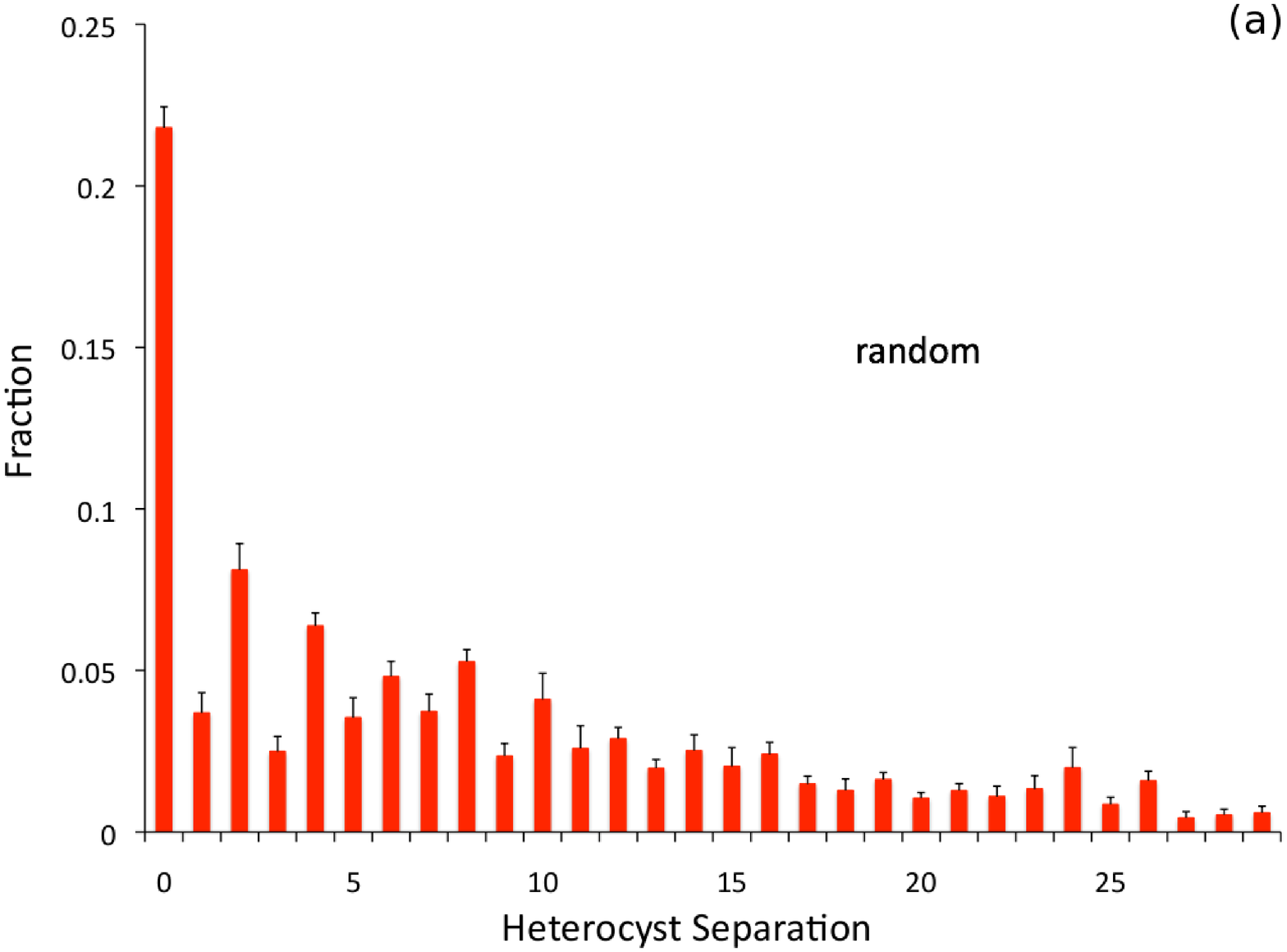}&\includegraphics[width=3.0in]{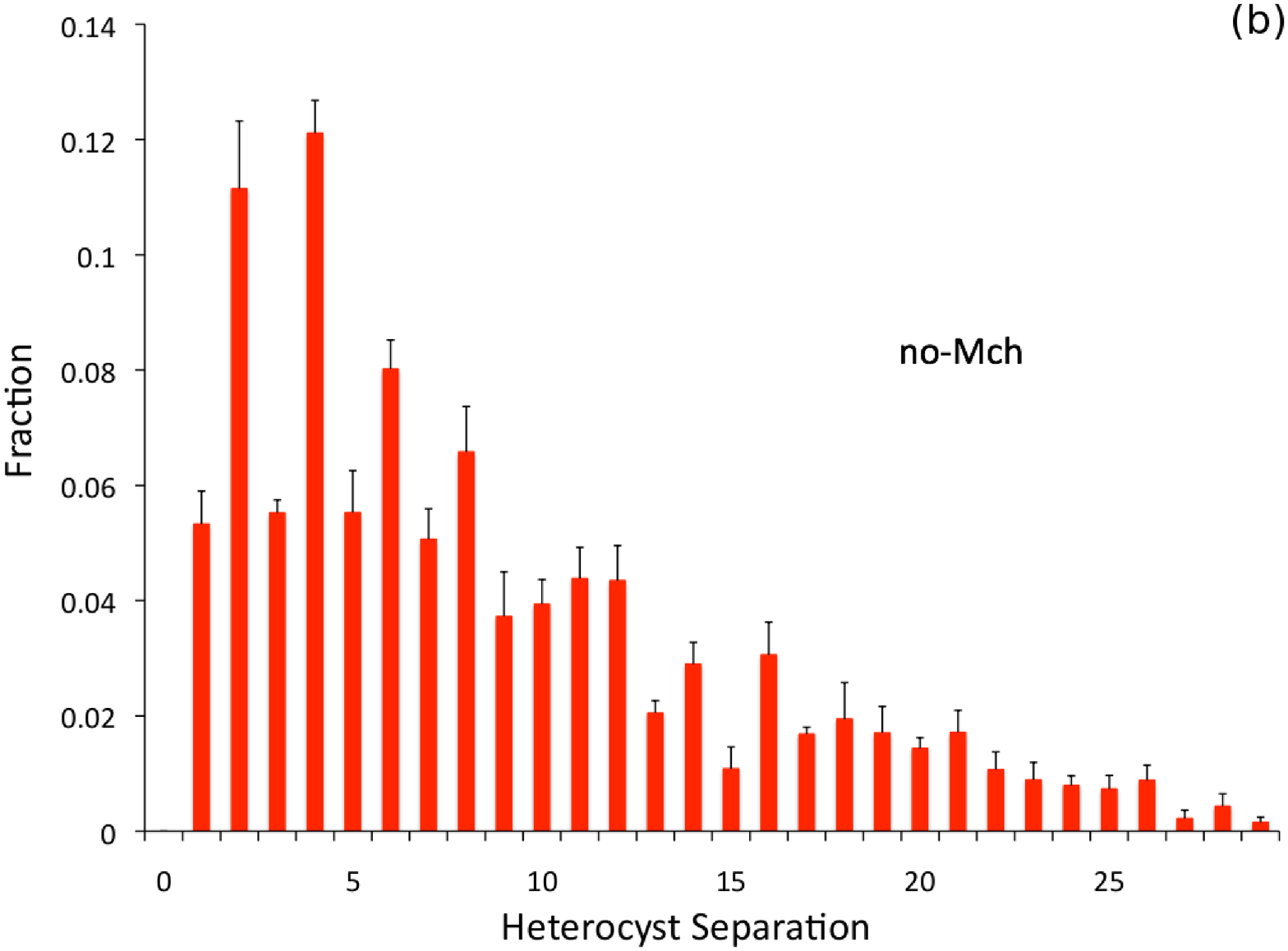} \\
    \includegraphics[width=3.0in]{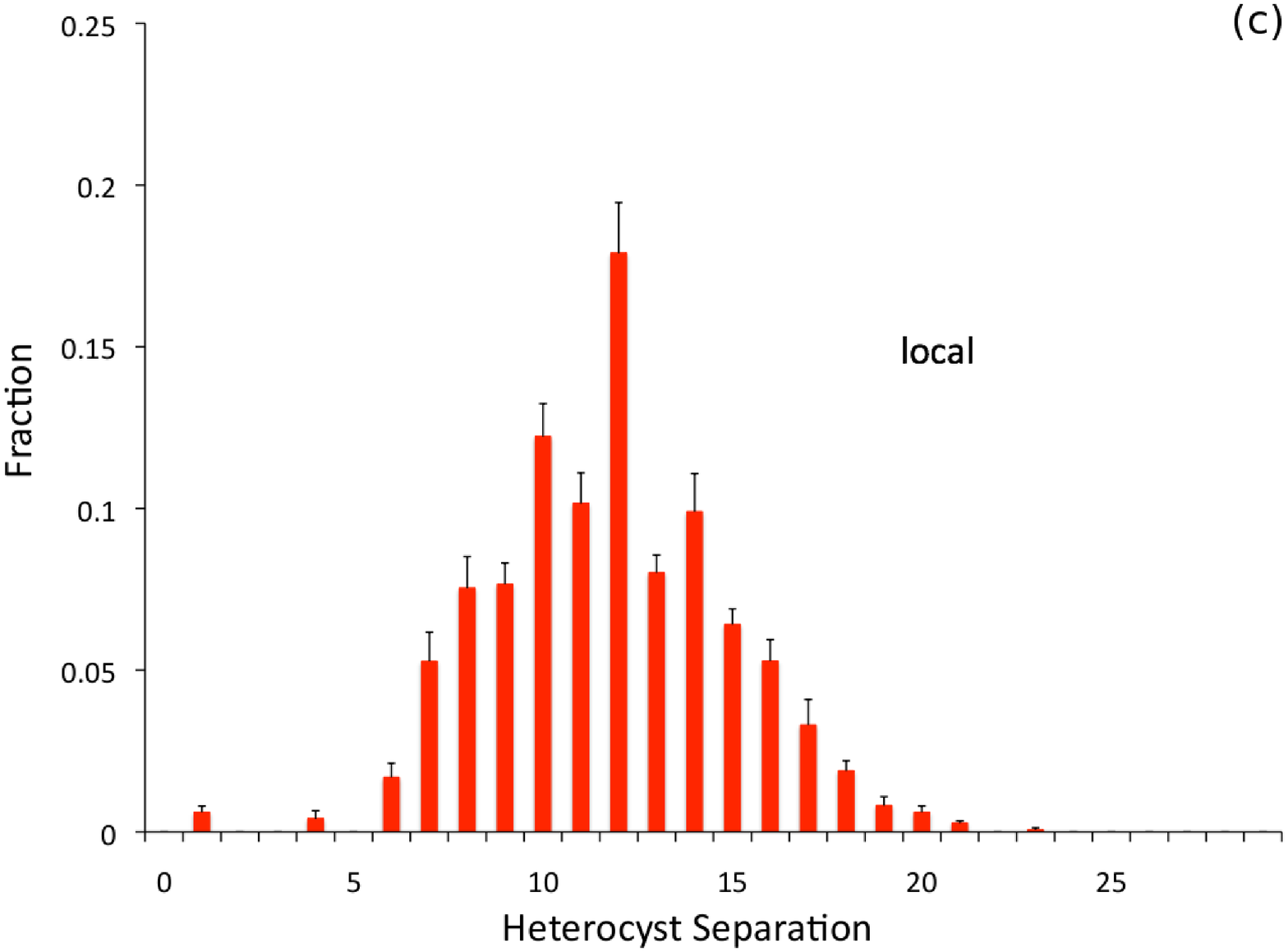}&\includegraphics[width=3.0in]{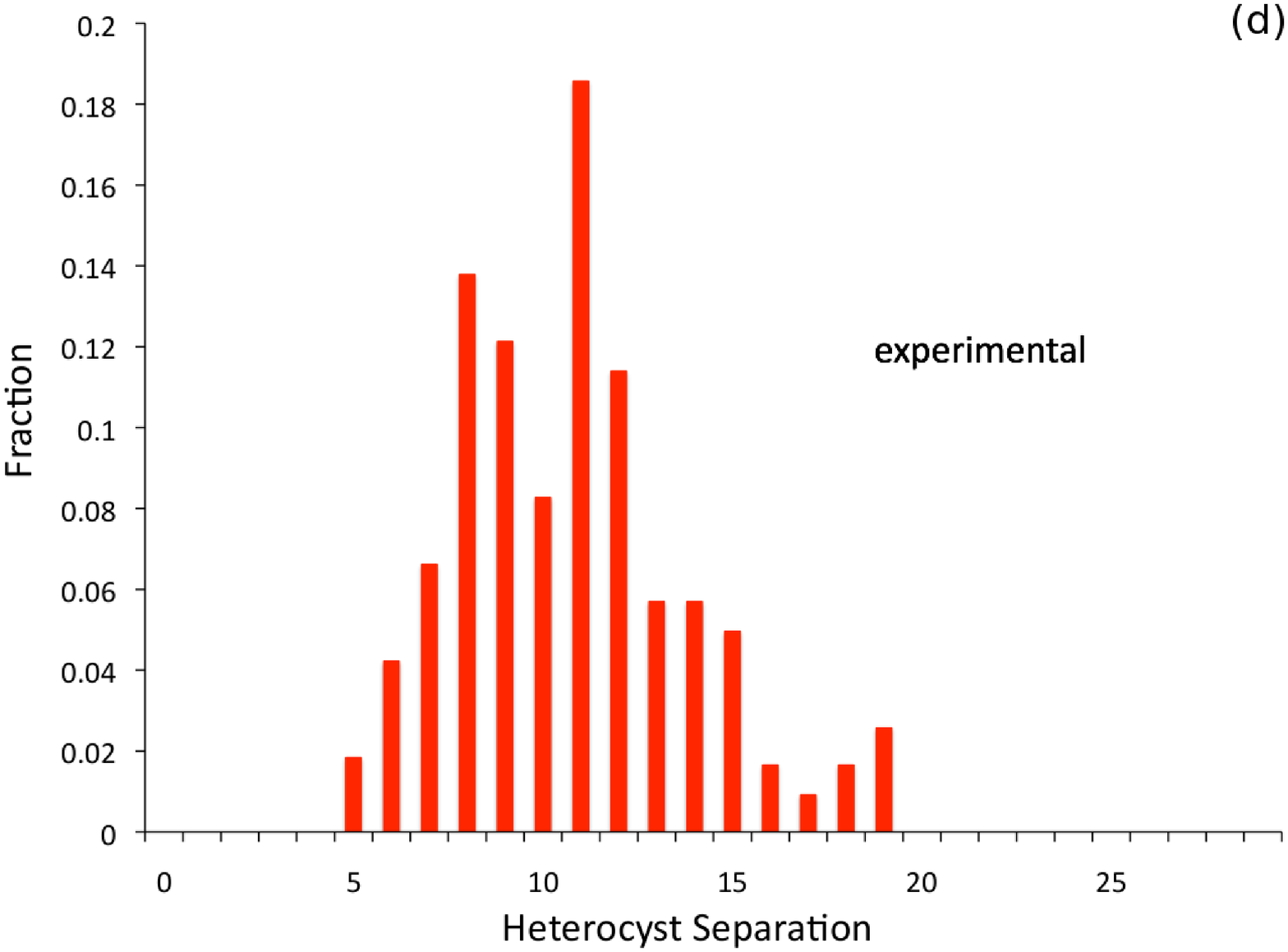}
  \end{tabular} 
  \end{center}
  \caption{\label{fig:sixth} (a)-(c) are heterocyst spacing distributions from stochastic simulations with 1$\%$ leakage and zero external fixed-nitrogen with error bars representing the standard deviation of the mean. (a) is random heterocyst placement with 10$\%$ heterocysts, (b) is no-Mch heterocyst placement with 10$\%$ heterocysts, and (c) is local heterocyst placement with period of starvation until commitment of $\tau=8$h. (d) is the experimental steady-state WT distribution from \cite{yoon01}.}
\end{figure}

Figs.~\ref{fig:sixth}(a)-(c) show heterocyst spacing distributions for the different heterocyst placement strategies with 1$\%$ leakage, all with approximately $f=0.1$ for ease of comparison. Both random placement and no-Mch favour small heterocyst spacings. Random placement, in Fig.~\ref{fig:sixth}(a), peaks at adjacent heterocysts (corresponding to the Mch phenotype) and drops off for larger separations, while no-Mch placement, in Fig.~\ref{fig:sixth}(b), peaks at a spacing of 4 intercalating cells between heterocysts. Significant bias towards even spacings, due to ongoing filament growth, is also seen. The distribution for local placement is quite different, with a symmetric peak at approximately 12 intercalating cells and very few heterocysts separated by less than 6 cells. Fig.~\ref{fig:sixth}(d) is an experimental spacing distribution after $96$h of fixed-nitrogen deprivation for WT \emph{Anabaena} PCC 7120 \cite{yoon01}. The heterocyst spacing distribution with the local placement strategy is qualitatively similar to the experimental distribution, even though the strategy selection was done with respect to growth alone --- without consideration of the spacing distribution. 

\begin{figure}[!ht]
 \begin{center}
  \begin{tabular}{cc}
    \includegraphics[width=3.0in]{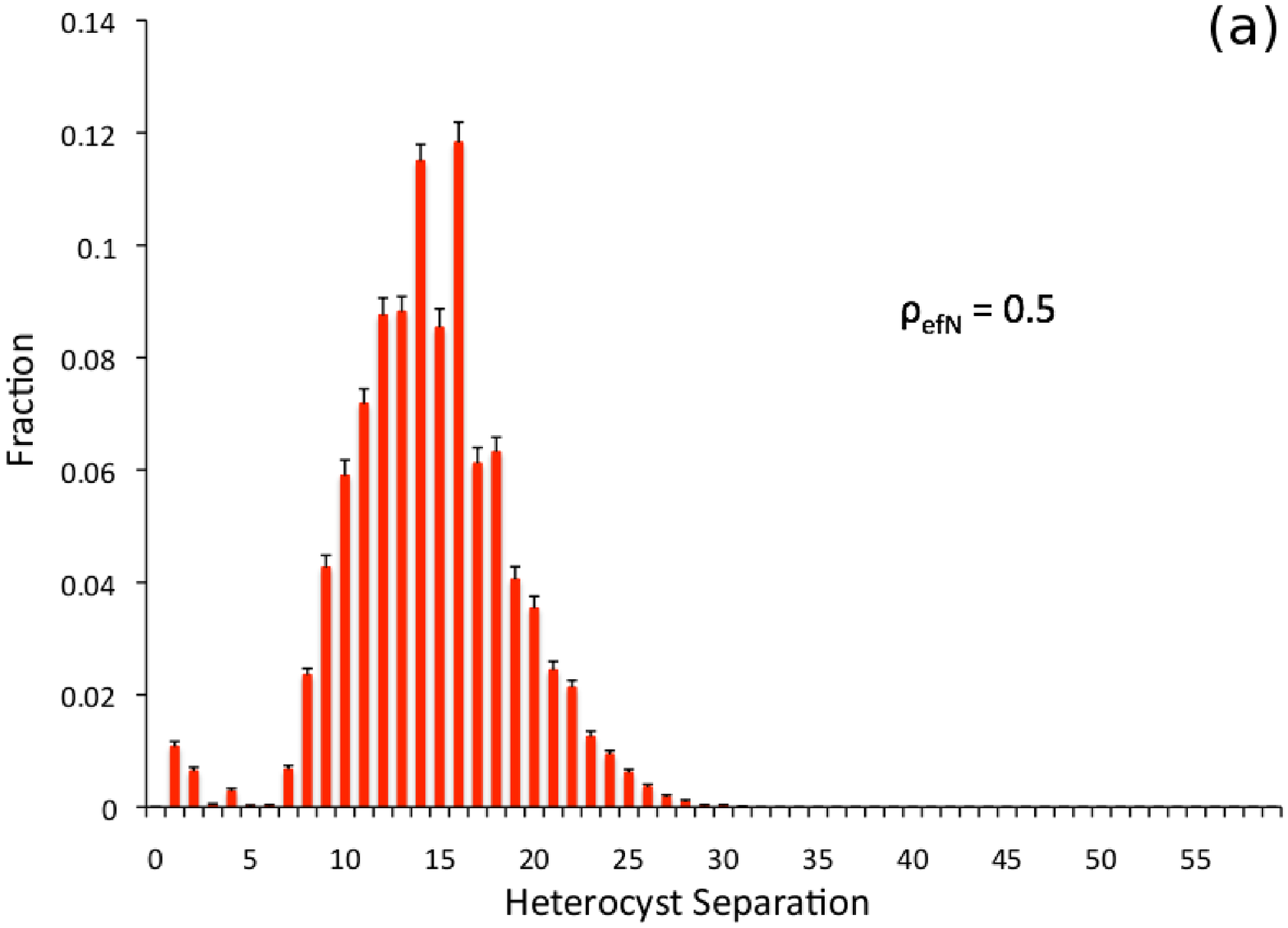}&\includegraphics[width=3.0in]{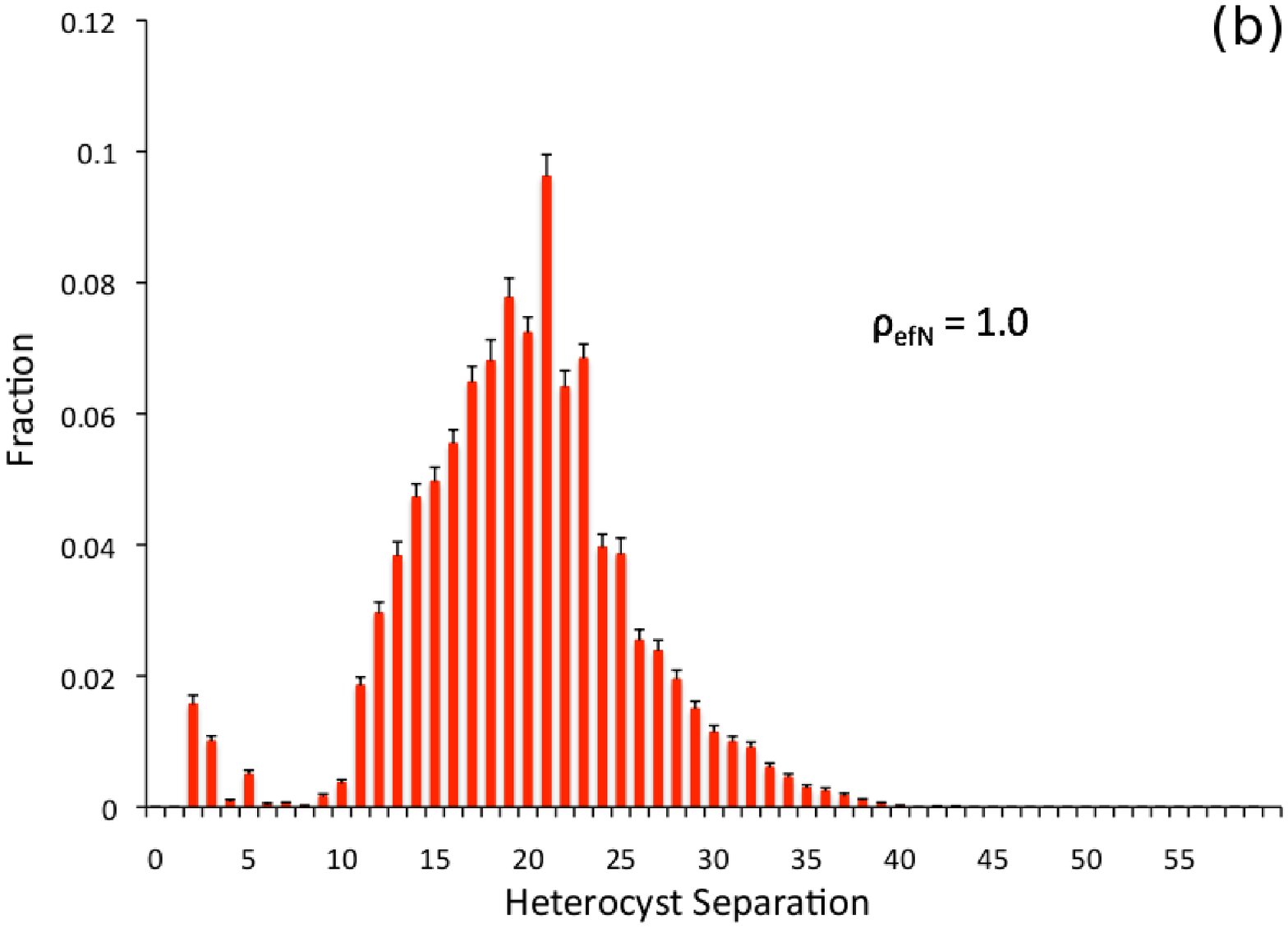} \\
    \includegraphics[width=3.0in]{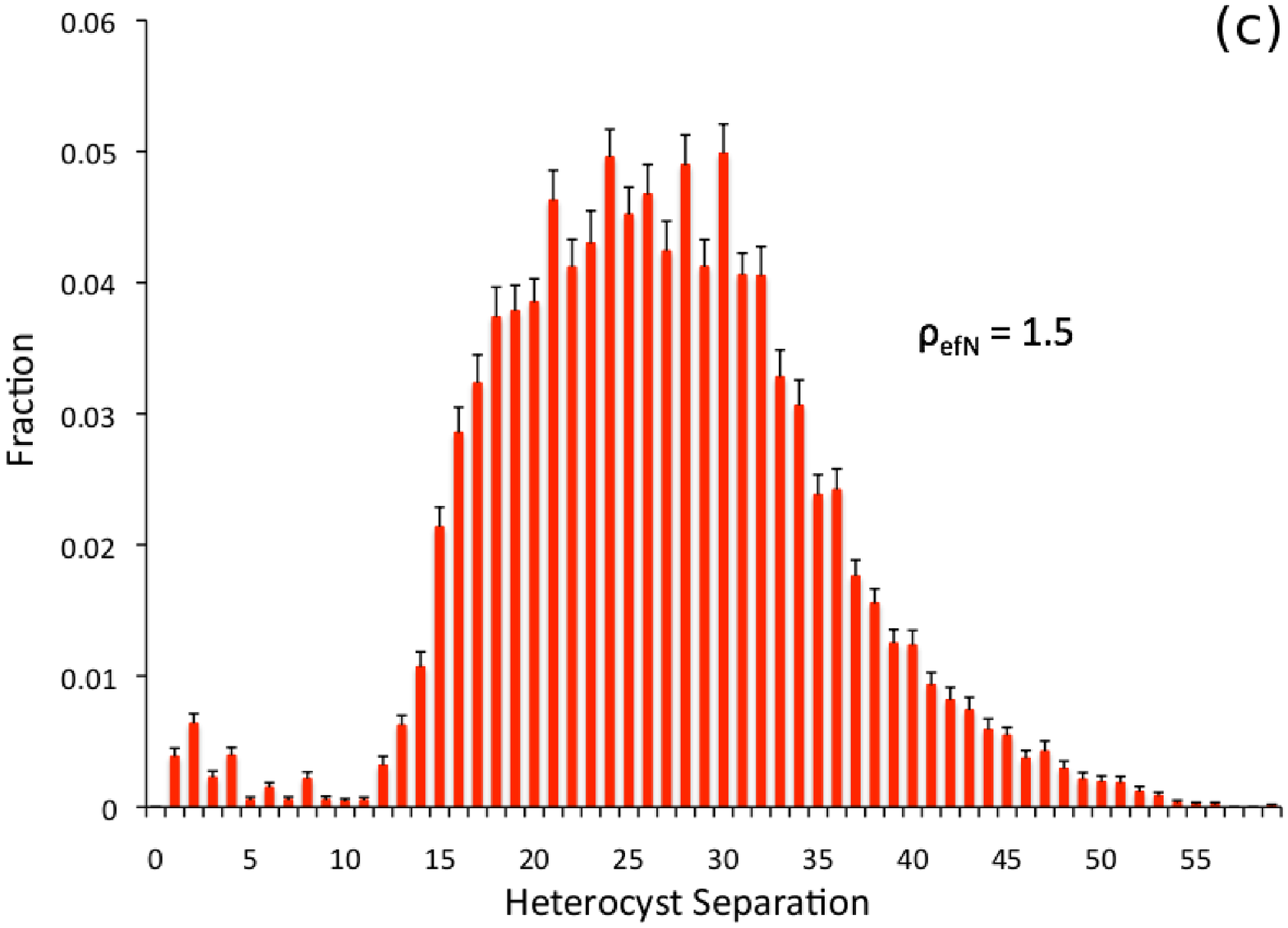}&\includegraphics[width=3.0in]{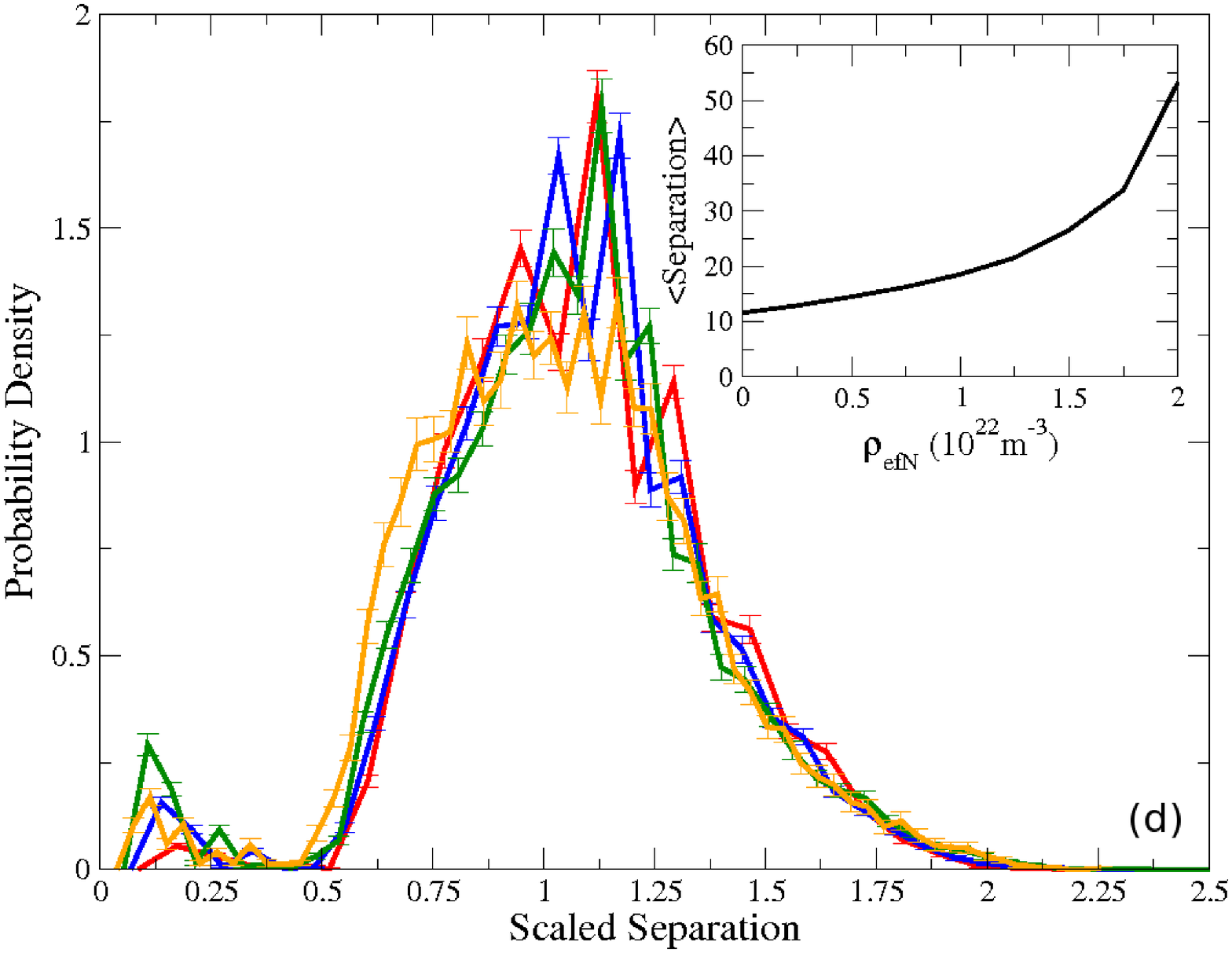}
  \end{tabular} 
  \end{center}
  \caption{\label{fig:seventh} Heterocyst spacing distributions for filaments with local heterocyst placement, 1$\%$ leakage for various $\rho_{efN}$  and a period of starvation until committment of $\tau=8$h. (a)  $\rho_{efN} = 0.5 \times 10^{22}$m$^{-3}$, (b) $\rho_{efN} =  1 \times 10^{22}$m$^{-3}$, and (c) $\rho_{efN}=   1.5 \times 10^{22}$m$^{-3}$. (d) shows the probability density of the distributions with $\rho_{efN}=$ 0 (red), 0.5 (blue), 1 (green), and 1.5 (orange) $\times$10$^{22}$m$^{-3}$ vs. scaled heterocyst separation. Inset in (d) is the average heterocyst separation vs. efN concentration.}
\end{figure}

With non-zero $\rho_{efN}$ we consider only the heterocyst spacing distributions for the local heterocyst placement strategy, as illustrated in  Figs.~\ref{fig:seventh}(a)-(c) with 1$\%$ leakage. The peak separation increases with $\rho_{efN}$, consistent with the decreasing $f^\ast$ we saw in Fig.~\ref{fig:fourth} (b).   The distribution also becomes significantly wider. By scaling the heterocyst separation by the average separation for each distribution we see  in Fig.~\ref{fig:seventh} (d) that the distributions at different $\rho_{efN}$ approximately collapse to a single scaled distribution independent of $\rho_{efN}$. The inset in Fig.~\ref{fig:seventh}(d) is the average heterocyst separation vs. $\rho_{efN}$. 

\section{Discussion}

We have explored three heterocyst placement strategies in a cyanobacterial filament with a quantitative model for fixed-nitrogen (fN) transport and dynamics that includes growth of vegetative cells and production of fN by heterocysts. The three strategies were random placement, random placement with no-Mch, and local placement.  For random strategies, we found an optimal heterocyst frequency $f^\ast$ at which the growth rate of the filament is maximized, $\mu^\ast$. With fewer heterocysts than $f^\ast$  filaments are fN limited. With more heterocysts than $f^\ast$ the filament is hampered by an excess of non-growing heterocysts.  As the external fN, $\rho_{efN}$, is increased from zero, the optimal heterocyst frequency decreases continuously until it reaches zero at a concentration $\rho_{efN}^\ast$ where all fN needs are met by imported extracellular fN. The local placement strategy led to a similar decreasing heterocyst fraction $f^\ast$ with $\rho_{efN}$, though without explicit optimization of the heterocyst frequency.

Without leakage of cytoplasmic fN from the filament, via $D_L$, the difference between these strategies is small (see Fig.~\ref{fig:third}, red lines and circles). With leakage there are significant growth differences between the strategies under limiting efN conditions (with $\rho_{efN}< \rho_{efN}^\ast$).  Filaments using local placement of heterocysts grow faster than filaments with no-Mch, which in turn grow faster than filaments using random heterocyst placement.   The differences between the strategies are largest when $\rho_{efN}=0$, and decrease as $\rho_{efN}$ increases --- until it vanishes above $\rho_{efN}^\ast$ where the heterocyst frequency $f^\ast=0$ for all three strategies.   We suggest that leakage of fN from growing filaments may be important for understanding the adaptive nature of heterocyst placement strategies.

We believe that the observed growth differences  are relevant. Selective sweeps occur when a beneficial mutation `sweeps' a population and becomes fixed \cite{imhof00}. Selective sweeps have been observed in cultures of \emph{E. coli} for mutations with a fitness parameter as small as $m=0.006$ \cite{imhof00}, which in our system corresponds to a growth rate constant difference of 0.0003h$^{-1}$.  Fig.~\ref{fig:fourth}(a) shows that the growth rate difference between random heterocyst placement strategies and local heterocyst placement is about 0.0025h$^{-1}$ for 1$\%$ leakage for zero external fixed-nitrogen, and is similar for 10$\%$ leakage --- substantially larger than necessary for a selective sweep in {\em E. coli}.  

Significantly, we did not pick our best strategy to match the heterocyst spacing pattern nor did we tweak the heterocyst frequency $f$. Rather we implemented a simple local strategy which itself chose $f^\ast$. This best, local strategy resulted in faster growth than optimized random strategies, and in heterocyst spacings remarkably similar to those seen experimentally.  The similarity of our model heterocyst spacing distributions with observed patterns suggests that local fN starvation may drive heterocyst development. Indeed, we believe that the extensive genetic network of {\em Anabaena} heterocyst development \cite{kumar10} effectively implements something similar to our local strategy.  This is consistent with  experimental work  \cite{popa07} (see also \cite{brown12}) showing a dip in the fixed-nitrogen level approximately halfway between two widely separated heterocysts.

Qualitatively our best strategy is easy to state: ``quickly differentiate cells that are locally starving of fixed-nitrogen''. Effectively implementing the strategy is not trivial. We note two complications. The first is that our local strategy works for ongoing differentiation of heterocysts, during steady-state growth. In that regime, long after efN deprivation, a dedicated mechanism to avoid Mch is not needed --- because nearby cells begin to starve at different times. However, in the first burst of differentiation we need to avoid Mch with a temporary mechanism to prevent differentiation of the entire filament. Indeed, the initial heterocyst spacing is distinct from the steady-state pattern \cite{yoon01} and cyanobacterial filaments are thought to use diffusible inhibitors derived from PatS to suppress Mch \cite{yoon98}. The second is that we have simplified heterocyst commitment \cite{yoon01} and subsequent delay before {\em du novo} nitrogen fixation into a single delay $\tau$.  While we find that a smaller $\tau$ always leads to faster filament growth, heterocysts do take significant time to both commit and to begin to fix nitrogen \cite{kumar10, yoon01, ehira03, thiel01} (though see \cite{allard07}).  While we have explored a correspondingly large range of $\tau \in \{ 1-20\}$h, with similar results throughout, the details may be expected to change with a more detailed model of heterocyst commitment. However our understanding of heterocyst commitment timing with respect to local fN concentrations remains crude, and we feel that a unified delay $\tau$ is an appropriate simplification.

We predict a plastic developmental response of heterocyst frequency to levels of efN, where the heterocyst frequency $f$  decreases rapidly with $\rho_{efN}$. This implicitly assumes that there is no threshold extracellular fixed-nitrogen concentration above which heterocysts will not differentiate, and more broadly that there is not a fixed developmental pattern of heterocysts. Supporting a plastic developmental response, early work by Fogg \cite{fogg49} showed a time-dependent heterocyst frequency that increased with decreasing levels of efN, Ogawa and Carr \cite{ogawa69} found heterocyst frequency and nitrate concentration to be inversely related, and Horne {\em et al} \cite{horne79} found the same correlation in field studies.  Fogg also observed significant heterocyst development at $40 \mu$M extracellular fixed-nitrogen (Fig.~2 of \cite{fogg49}), though not in steady-state conditions. Plastic developmental responses are observed in systems ranging from plants \cite{sultan00} to the brain \cite{holtmaat09}. We feel that filamentous cyanobacteria are particularly amenable to exploring and understanding the adaptiveness of this plastic response. The time is now ripe to undertake constant external fixed-nitrogen concentrations with single-filament studies in the chemostat-like environment of microfluidic systems, such as filamentous cyanobacteria within the long channels used in studies of persister cells of {\em E. coli} \cite{balaban04, bennett09}.  

We note that the heterocyst spacing patterns themselves are not adaptive. Rather the patterns reflect an adaptive local strategy for placing heterocysts close to locally starving cells.  This starvation is a combination of distance from heterocysts and fast growing cells. Local placement is particularly adaptive when fN leaks from the cyanobacterial filament (see \cite{fogg49, walsby75, paerl78, thiel90, pernil08, picossi05}, as discussed in the introduction).  

Our heterocyst placement strategy is deliberately simple, and does not include any diffusible inhibitors such as peptides derived from PatS or HetN \cite{yoon98, wu04, risser09}. We anticipate that the small secondary peak that we observe at smaller separations in Figs.~\ref{fig:seventh} (a)-(d) may be suppressed \emph{in vivo} by the action of such diffusible inhibitors. Since HetN appears to act after initial heterocyst development \cite{callahan01}, it may indeed be why the secondary peak is not actually observed \cite{yoon01}. However, we note that the separation corresponding to the secondary peak increases with the average separation, so that the secondary peak may emerge in experimental studies at larger levels of $\rho_{efN}$ if inhibition due to PatS and HetN has a fixed range.

Our model and its results assume that, apart from fixed-nitrogen, there are no nutrient or metabolite requirements that limit growth differentially along the filament. Doubtless, there are experimental conditions where this assumption is not warranted. Certainly, there are interesting nutrients that we have not included in our model. For example, the supply of membrane potential and/or of carbohydrates from vegetative cells to heterocysts \cite{turpin85, cumino07} was not considered.  Nevertheless, while carbohydrate transport would be expected to starve the interior cells of clusters of  contiguous heterocysts and reduce their capacity for nitrogen-fixation, our primary focus was on the local heterocyst placement strategy under conditions of non-zero external fixed-nitrogen --- where no clusters of heterocysts are observed and heterocysts are broadly spaced.  However, carbohydrate limitation would affect random placement strategies, which might then have even more reduced growth compared to our best, local strategy. 

For cyanobacterial filaments implementing a local heterocyst placement strategy we have several experimentally testable predictions that we expect to be observed in steady-state conditions --- independent of the specific parameterization of our quantitative model. To explore them in the lab, external concentrations of fixed-nitrogen ($\rho_{efN}$) would need to be controlled by microfluidic devices \cite{balaban04, bennett09} or flow cells. The first is that heterocyst frequency $f^\ast$ will decrease rapidly and continuously as $\rho_{efN}$ is increased from zero. As shown in Figs.~\ref{fig:fourth} (b) and (d), the dependence is markedly non-linear with larger leakage rates of fN from the growing filament. The second, related, prediction is that the peak of the heterocyst spacing distribution will increase as $\rho_{efN}$ decreases. This is illustrated in Figs.~\ref{fig:seventh} (a)-(c).  We also expect that the width, or standard deviation, of the heterocyst spacing will increase as the mean spacing increases. Indeed, for our model, the standard deviation and mean spacing are approximately proportional.

Our results may also have implications for quantitative models of biogeochemical cycling of fixed-nitrogen in marine and lake environments, where filamentous cyanobacteria can play a significant role. We note that a number of existing models \cite{tyrrell99,bissett99,neumann00} use growth rates and nitrogen-fixation rates that are independent of the biologically available fixed-nitrogen. We  find that, with a local heterocyst placement strategy, the filament growth rate $\mu^\ast$ is approximately (but not exactly) independent of $\rho_{efN}$ --- see Fig.~\ref{fig:fourth} (a) and (c), consistent with the approximately constant growth reported for chemostat studies by Elder and Parker \cite{elder84}. However we expect that {\em du nuovo} nitrogen-fixation to be proportional to $f^\ast$, which in turn strongly decreases with increasing $\rho_{efN}$. 

\ack
We thank the Natural Science and Engineering Research Council (NSERC) for support, and the Atlantic Computational Excellence Network (ACEnet) for computational resources. AIB also thanks NSERC, ACEnet, and the Killam Trusts for fellowship support.

\pagebreak

\section*{References}
\bibliographystyle{unsrt}
\bibliography{brownrutenberg7may2012}

\end{document}